\documentclass{amsproc}

\theoremstyle{definition}

\theoremstyle{remark}

\numberwithin{equation}{section}

\def\su#1{{\rm SU}( #1)}
\def\sp#1{{\rm Sp}( #1)}

\def\even{{\rm even}}
\def\odd{{\rm odd}}
\def\half{\frac{1}{2}}
\def\tshalf{{\textstyle \frac{1}{2}}}
\def\fourth{\frac{1}{4}}
\def\threehalf{\frac{3}{2}}
\def\threefourth{\frac{3}{4}}

\newcommand{\Z}{\mathbb{Z}}
\newcommand{\R}{\mathbb{R}}
\newcommand{\C}{\mathbb{C}}

\newcommand{\N}{\mathcal{N}}

\def\sqr#1#2{{\vcenter{\vbox{\hrule height.#2pt
    \hbox{\vrule width.#2pt height#1pt \kern#1pt
    \vrule width.#2pt}
    \hrule height.#2pt}}}}

\newdimen\tableauside\tableauside=1.0ex
\newdimen\tableaurule\tableaurule=0.4pt
\newdimen\tableaustep
\def\phantomhrule#1{\hbox{\vbox to0pt{\hrule height\tableaurule width#1\vss}}}
\def\phantomvrule#1{\vbox{\hbox to0pt{\vrule width\tableaurule height#1\hss}}}
\def\sqr{\vbox{%
\phantomhrule\tableaustep

\hbox{\phantomvrule\tableaustep\kern\tableaustep\phantomvrule\tableaustep}%
  \hbox{\vbox{\phantomhrule\tableauside}\kern-\tableaurule}}}
\def\squares#1{\hbox{\count0=#1\noindent\loop\sqr
  \advance\count0 by-1 \ifnum\count0>0\repeat}}
\def\tableau#1{\vcenter{\offinterlineskip
  \tableaustep=\tableauside\advance\tableaustep by-\tableaurule
  \kern\normallineskip\hbox
    {\kern\normallineskip\vbox
      {\gettableau#1 0 }%
     \kern\normallineskip\kern\tableaurule}%
  \kern\normallineskip\kern\tableaurule}}
\def\gettableau#1 {\ifnum#1=0\let\next=\null\else
  \squares{#1}\let\next=\gettableau\fi\next}
\newcommand{\Yfund}{\tableau{1}}

\newcommand{\Yasymm}{\tableau{1 1}}

\usepackage{graphics}
\usepackage{epsfig}

\begin{document}
\title{Elliptic Fibrations and Elliptic Models}

\author{Amy E. Ksir}
\address{Department of Mathematics, State University of New York at
Stony Brook, Stony Brook, NY 11794}
\email{ksir@math.sunysb.edu}
\thanks{The first author was supported in part by NSF Grant DMS-9983196 
as a VIGRE Postdoctoral Fellow.}

\author{Stephen G. Naculich}
\address{Department of Physics, Bowdoin College, Brunswick, ME 04011}
\email{naculich@bowdoin.edu}
\thanks{The second author was supported in part by NSF Grant
PHY-9407194 through the ITP Scholars Program.}

\date{March 27, 2002}


\begin{abstract}
    We study the Seiberg-Witten curves for $\N=2$ SUSY gauge theories
    arising from type IIA string configurations with two orientifold sixplanes.
    Such theories lift to elliptic models in M-theory.
    We express the M-theory background for these models as
    a nontrivial elliptic fibration over $\C$.
    We discuss singularities of this surface,
    and write the Seiberg-Witten curve for several theories
    as a subvariety of this surface.
\end{abstract}

\maketitle

\section{Introduction}

In the Seiberg-Witten approach to $\N =2$ supersymmetric gauge
theory \cite{SW}, one identifies a family of algebraic curves associated to
each choice of a gauge group and matter content.  One approach to this
is via M-theory \cite{W4D}.  One identifies a configuration of branes in type
IIA string theory such that the induced theory on the world
volume of the D4 branes is the desired gauge theory.   The D4 branes and
NS5 branes in the IIA theory then lift to an M-theory five brane,
whose world volume is $\R^{4} \times \Sigma$,
where $\Sigma$ is the
desired algebraic curve.  This curve is embedded in an algebraic
surface $Q$, where $\R^{7} \times Q$ is the eleven-dimensional
M-theory background.
(For brevity, we will refer to $Q$ as the M-theory background.)

In some cases, the type IIA configuration contains D6 branes and
orientifold six planes as well as the NS5 and D4 branes.
These affect the geometry of $Q$.
In cases with D6 branes and no orientifold planes,
$Q$ is a multi-Taub-NUT space \cite{W4D}.
In cases with one (negatively-charged) orientifold plane,
$Q$ is the Atiyah-Hitchin monopole moduli space \cite{LL,LLL,AH}.
In cases with two orientifold planes,
$Q$ is an elliptic surface, and the M-theory model is
an elliptic model.  This is the case that we study in this paper.

For elliptic models without orientifold planes,
the M-theory background $Q$ is of the form $\R^2 \times T^2$,
 which can be given the complex structure $\C \times E$,
where $E$ is an elliptic curve.
More generally, the background can be
an affine $\C$ bundle over $E$.
The Seiberg-Witten curve is then written as a cover over $E$.
For elliptic models with orientifold planes,
in cases where the orientifold plane charge is cancelled
{\it locally} by D6 branes \cite{U,GK},
the M-theory background can be viewed
as the quotient of an affine $\C$ bundle over $E$
by a $\Z_2$ action.
It is not clear, however, how to
extend this to
more general orientifold backgrounds.

In this paper, we adopt a different approach,
viewing the M-theory background $Q$ as an elliptic fibration over $\C$,
and the Seiberg-Witten curve as a cover of $\C$.
This approach is partially motivated by the results of ref.~\cite{2anti},
in which the Seiberg-Witten curves were written
in terms of theta functions with a varying modular parameter.
In section 2, we give the explicit form of the elliptic surface $Q$
for the background corresponding to two negatively-charged
orientifold six-planes with coincident D6 branes.
We briefly discuss the singularities of this surface in the context of
M-theory and F-theory.
In section 3, we derive the explicit form of the Seiberg-Witten curves
for three $\N=2$ gauge theories with this background:
$\sp{2k}$ + 1 antisymmetric + 4 fundamental hypermultiplets,
$\sp{2k}\times\sp{2k}$ + 1 bifundamental + 4 fundamental
hypermultiplets, and $\su{N}$ + 2 antisymmetric + 4 fundamental
hypermultiplets.  We show that our results are in agreement with the curves
for these theories derived using different approaches \cite{U,GK,ELNS}.

We expect that the description of the M-theory background $Q$
as an elliptic fibration will generalize to the situation
where the D6 branes are displaced from the orientifold sixplanes,
by a deformation of the elliptic fibration,
as occurs in F-theory \cite{Sen}.
Knowing the precise form of the M-theory background
for these more general brane configurations
is crucial to determining the Seiberg-Witten curve
for theories in which the fundamental hypermultiplets have
nonzero masses, particularly the terms of the curve equation
subleading in the QCD scale $\Lambda$,
which are currently unknown \cite{ELNS}.

\section{The surface}

We consider the M-theory background corresponding to type IIA models with two
orientifold six planes.
An orientifold six plane can have charge $+4$ or $-4$ relative to
the D6-brane charge; we consider models with two negatively-charged
O$6^{-}$ planes, and add four pairs of D6 branes to the theory to cancel
the charge.
Each orientifold six plane and D6 brane is extended
in the 0123789 directions.
We combine the $x_{4}$ and $x_{5}$ coordinates into the complex
coordinate $v=x_{4}+ix_{5}$, and place one of the orientifold planes at
$(v, x_{6})=(0,0)$
and the other at $(v, x_{6})=(m, \pi L)$,
where $m$ is the global mass \cite{U}.
Each orientifold plane is the fixed point set of a reflection,
and the two reflections
generate a translation symmetry $(v,x_6) \to (v+2m, x_6 + 2\pi L)$
of the IIA model.

The corresponding M-theory background,
which is invariant under $x_{10} \to x_{10} + 2\pi R$,
is thus doubly periodic, and gives rise to elliptic models.
The metric in the $x_{6}$ and $x_{10}$ directions is not generally a simple
product $S^{1} \times S^{1}$, but is such that travelling around the
$x_{6}$ direction results in a shift in the $x_{10}$ direction by $\theta R$.
The orientifold reflections lift to
\begin{eqnarray}
(v, x_{6}, x_{10}) &\to& (-v, -x_{6}, -x_{10})
\nonumber\\
(v,  x_{6}, x_{10}) &\to& (2m - v, 2\pi L - x_{6},
\theta R -x_{10}),
\end{eqnarray}
which have four fixed points
\begin{equation}
\label{four fixed points}
(v,x_{6},x_{10})= (0,0,0), \quad
(0,0,\pi R), \quad
(m, -\pi L, -\tshalf \theta R), \quad
(m, -\pi L, (\pi - \tshalf \theta) R).
\end{equation}
The fundamental parallelogram, with the four fixed points,
is shown in fig.~1.

\begin{figure}[t]
\begin{center}
\vbox{\epsfig{figure=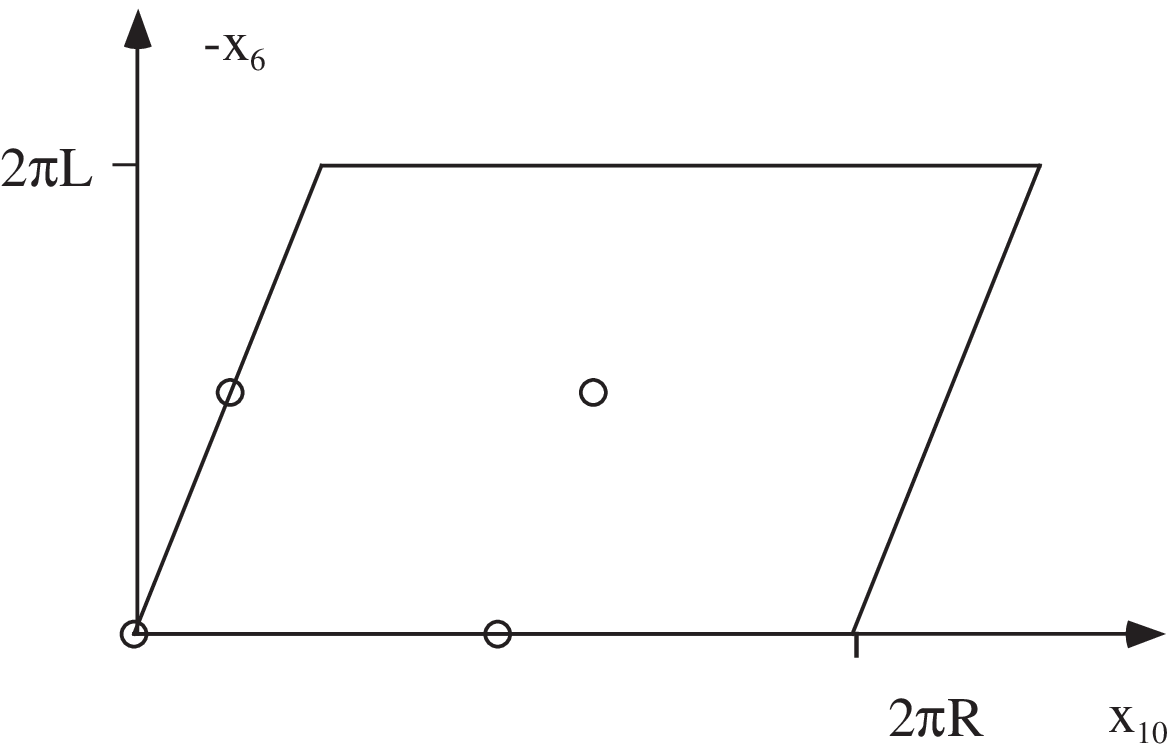, scale=.5}}
\end{center}
\begin{center}
{\footnotesize{\bf Fig. 1: A fundamental parallelogram for the
elliptic model. \\ Circles indicate the fixed points of the orientifold
involution.}}
\end{center}
\end{figure}

Now we consider the case where the global mass $m$ vanishes.
Defining
\begin{equation}
\nu=\frac{x_{6}+ix_{10}}{2\pi iR},
\end{equation}
the background is therefore invariant under
$\nu \to \nu +1$ and $\nu \to \nu+\tau$,
where
$\tau = - \frac{\theta}{2\pi} + i \frac{L}{R}$.
This gives $\R^2 \times T^2$ the complex structure of $\C \times E$,
where $\C$ is the $v$-plane
and $E$ is the quotient of the $\nu$-plane by the lattice $\Z \oplus \Z \tau$.
We can now embed the elliptic curve $E$ as a cubic in $\C P^{2}$ with local
equation
\begin{equation}
\label{cubic}
y^{2} = (x^{2}-4)(x-\lambda).
\end{equation}

If in addition the D6 branes are coincident with the O$6^-$ planes,
the M-theory background $Q$ is precisely the $\Z_{2}$ quotient of the
product $\C \times E$,
where the $\Z_2$-involution sends $(v,x,y) \to (-v,x,-y)$
in the cubic equation (\ref{cubic}).
We now express this quotient as an elliptic fibration.
Define invariant variables $u=v^2$ and $\eta = y v$.
Then $Q$ is characterized locally in $\C^{3}$ by the equation
\begin{equation}
\label{fibration}
\eta^2 = u (x^2-4)(x-\lambda),
\end{equation}
an elliptic fibration over the complex $u$-plane.
For nonzero values of $u$, the fiber is isomorphic (via rescaling) to
the original cubic curve $E$, so the parameter $\tau$ is constant
for $u \neq 0$.  At $u=0$, the fiber is singular and
consists of the line $u=\eta=0$,
with three singular points at $x=2,-2, \lambda$,
plus a line at infinity.
The local equation for each singular point is
$\eta^{2} \sim u (x-e_i)$,
where $e_i=2,-2, \lambda$, so these are
$A_{1}$ type singularities.  If we blow down the line at infinity, we get
one more $A_{1}$ singularity on this fiber at $x=\infty$.

The elliptic fibration (\ref{fibration})
is somewhat analogous to the one that
arises in the F-theory background
corresponding to an O7 plane and coincident D7 branes
in type IIB string theory \cite{Sen}.
However, there is an important difference.
The F-theory fibration contains a $D_4$ type singularity,
whereas the M-theory fibration above
contains four distinct $A_1$ type singularities
on the singular fiber at $u=0$.
This difference reflects the fact that, in F-theory,
the value of $\tau$ at each fiber is the dilaton-axion
modulus, but coordinates along the fiber have
no physical meaning whereas, in M-theory, the coordinates along the fiber
correspond to the $x_6$ and $x_{10}$ coordinates
(as discussed in sec. 4.3 of ref. \cite{Toroidal}).
If the $D_{4}$ singularity of the F-theory model is resolved, the
fiber is a chain of five genus zero curves arranged so that their
dual graph is the affine $\tilde{D_{4}}$ Dynkin diagram.  If the
four curves on the ends are contracted, the result is a genus
zero curve with four $A_{1}$ singularities, which is the fiber of the
M-theory model.  This is very similar to the phenomenon discussed in
refs.~\cite{Sen, Toroidal}.

We can perform a check on the fibration (\ref{fibration})
by considering the limit $\tau \to i\infty$, causing $\lambda \to \infty$.
This limit corresponds to sending
one of the orientifold planes, and its two accompanying pairs of D6 branes,
to infinity.
The resulting M-theory background,
which corresponds to one orientifold plane
and two pairs of coincident D6 branes,
was shown in ref.~\cite{LLL}
to be the ``$D_2$'' surface $a^2 + b^2 z = 4 z$
(which has a pair of $A_1$ singularities).
By rescaling the surface (\ref{fibration}) to
\begin{equation}
 \eta^{2}=u(x^{2}-4)(-\frac{x}{\lambda} +1)
\end{equation}
and taking the limit $\lambda \to \infty$, we obtain
$\eta^{2}= u(x^{2}-4)$,
which is identical to the surface above,
with $a = \eta$, $b = x$, and $z = -u$.

\section{The curves}

We will consider three $\N=2$ supersymmetric gauge theories arising
from type IIA brane configurations with two O$6^{-}$ planes,
with global mass $m=0$ and coincident D6 branes.
In each case, the IIA configuration contains
parallel NS5 branes, extended in the
012345 directions and separated in the $x_{6}$ direction,
and D4 branes extended in the 01236 directions, ending
on the NS5 branes in the $x_{6}$ direction.
Brane configurations for the three theories under consideration are
shown in fig.~2.

\begin{figure}[h]
\begin{center}
\vbox{\epsfig{figure=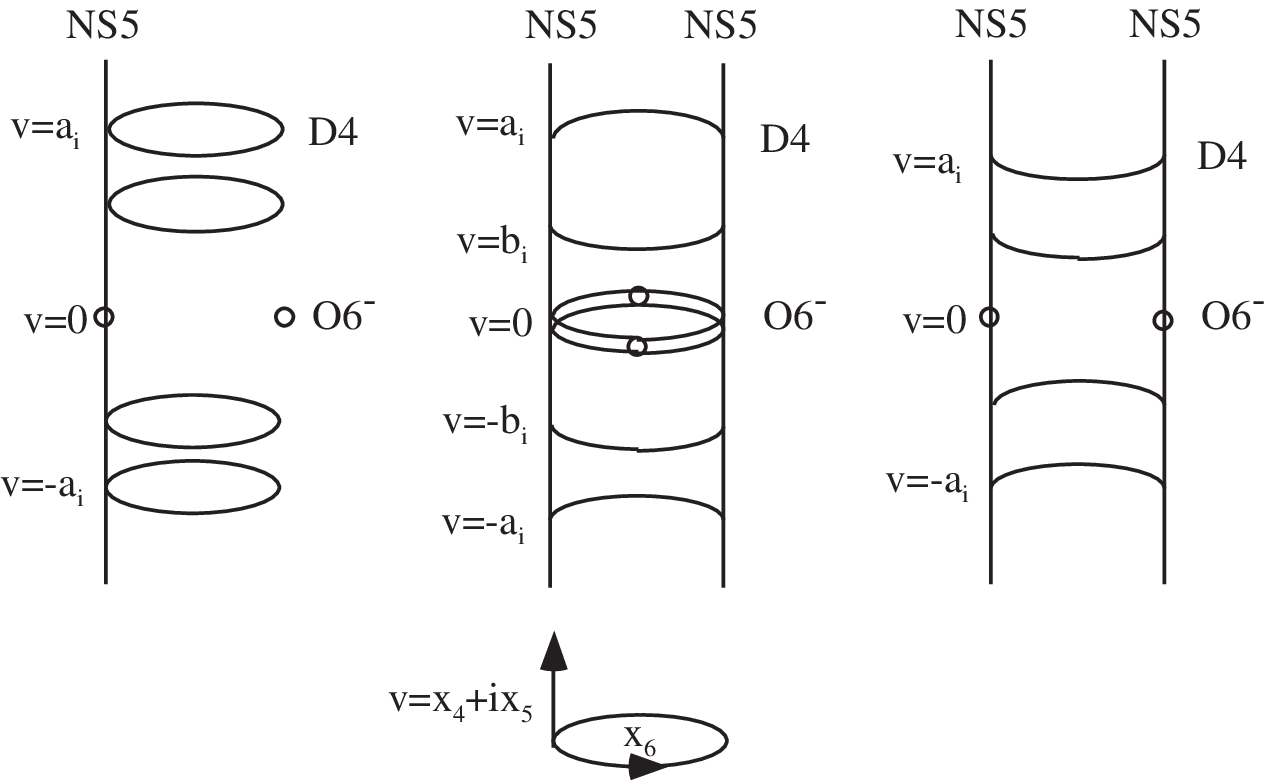, scale=.8}}
\end{center}
\begin{center}
{\footnotesize{\bf Fig. 2: Brane configurations for the three
theories. \\ Circles denote the positions of the O$6^{-}$ planes.}}
\end{center}
\end{figure}

For each theory, the IIA brane configuration lifts to
an M5-brane, whose embedding in $\R^{7} \times Q$ is given by
$\R^{4} \times \Sigma$, where $\Sigma$ is the Seiberg-Witten curve.
In the cases with two NS5 branes, we will express $\Sigma$ as a
double cover of the $u$-plane.
The NS5 branes correspond to the sheets of the cover,
and each D4 brane becomes a ``tube''
connecting the two sheets of the cover.
More precisely, each D4 brane (and its orientifold mirror)
corresponds to a branch cut in the $u$ plane.

This agrees with what we know from nonelliptic models, where a pair of
NS5 branes connected by a pair of D4-branes (at $v=\pm a$)
is represented by
$$t^2 + (u-a^2)t + 1= 0, $$
where $t =\exp [-(x_6 + i x_{10})/R]$.
The branch points occur at $u=a^2 \pm 2$, $t=\pm 1$,
i.e.,  $(x_6,x_{10}) = (0,0)$ and $(0,\pi R)$.
The branch cut in the $u$ plane, parallel to the real axis from
$a^2-2$ to $a^2+2$,
corresponds to $|t|=1$, i.e.,
the $x_{10}$ circle at the fixed value $x_6=0$ where the NS5 branes join.
The ``position'' of the D4-brane in the IIA picture, viz. $u=a^2$,
corresponds to $t = \pm i$, i.e.
points on the $x_{10}$ circle midway between the pre-images
of the branch points.

For each of the gauge theories in this section,
we will give an equation $F(u, x, \eta)=0$ for $\Sigma$ as a curve in
the surface $Q$ (\ref{fibration}).

\subsection{$\sp{2k}$ + 1 antisymmetric + 4 fundamental hypermultiplets}

The IIA configuration giving rise to the
$\N=2$ Sp($2k$) gauge theory with hypermultiplets in the
$\Yasymm + 4 \Yfund$ representations \cite{U}
contains only one NS5 brane,
intersecting the O$6^{-}$ plane at $u=\eta=0, x=2$.
Each physical D4 brane, located at $u =  a^2_{i}$
(corresponding to a mirror pair at $v=\pm a_i$),
wraps around
the $x_{6}$ direction and comes back to the same point on the NS5 brane.
In the M-theory lift,
the brane wraps the $x_{10}$ direction as well, so it is represented
by the entire fiber torus at $u=a_i^2$.
The NS5 brane is represented by $x=2$.
Thus the equation for the Seiberg-Witten curve is given by
\begin{equation}
    (x-2) \prod_{i=1}^k (u-a^2_{i})=0
\end{equation}
within the surface $Q$ given by equation (\ref{fibration}).
It is a reducible curve made up of the $x=2$
line along with the fiber tori at
each $u=a_{i}^2$.
This agrees with the results obtained in refs.~\cite{U} and \cite{ELNS}.

\subsection{$\sp{2k}\times\sp{2k}$ + 1 bifundamental + 4 fundamental
hypermultiplets}

The IIA configuration corresponding to
the $\N=2$ Sp($2k$) $\times$ Sp($2k$) gauge theory with hypermultiplets
in the $(\Yfund,\Yfund) + 2 (\Yfund,1) + 2 (1, \Yfund)$ representations
consists of two NS5 branes between the orientifold planes,
and a total of $2k$ physical D4 branes stretched between them \cite{U}.
Half of the D4 branes go around the $x_{6}$ circle in one
direction and the other half go in the other direction.
The two NS5 branes
are symmetric with respect to the orientifold $\Z_{2}$ action.
Furthermore,
we can choose them to be symmetric under the reflection
$(u,x,\eta) \to (u,x,-\eta)$.
The Seiberg-Witten curve can then be written as a
double cover of the $u$-plane by giving an equation
\begin{equation}
\label{doublecover}
x=f(u),
\end{equation}
with $f(u)$ to be determined.
To simplify calculations, we shift $x$
by a fractional linear transformation so
that the fibration locally has the form
\begin{equation}
\label{newfibration}
\eta^2 = u (x^2-4)(x^2-\mu^2).
\end{equation}
The two sheets of the cover correspond to
$(u,x,\eta)$ and $(u,x,-\eta)$ satisfying (\ref{doublecover})
and (\ref{newfibration}).
The branch points of the cover
therefore occur at points $u$ where $f(u)=2$, $-2$, $\mu$, or $-\mu$.

To determine the form of $f(u)$,
consider the picture on the universal cover,
$\C_{v} \times \C_{\nu}$ of the surface $Q$,
where $\nu=(x_{10}-ix_{6})/2\pi R$.
Choose $x$ so that $x=2$, $-2$, $\mu$, and  $-\mu$
correspond to $\nu = 0,$  $\half$,  $\frac{\tau}{2}$,
and $\half+\frac{\tau}{2}$ respectively (see fig.~3).
For each value of $v$,
the two NS5 brane positions are at $\nu$ and $-\nu$.

\begin{figure}[t]
\begin{center}
\vbox{\epsfig{figure=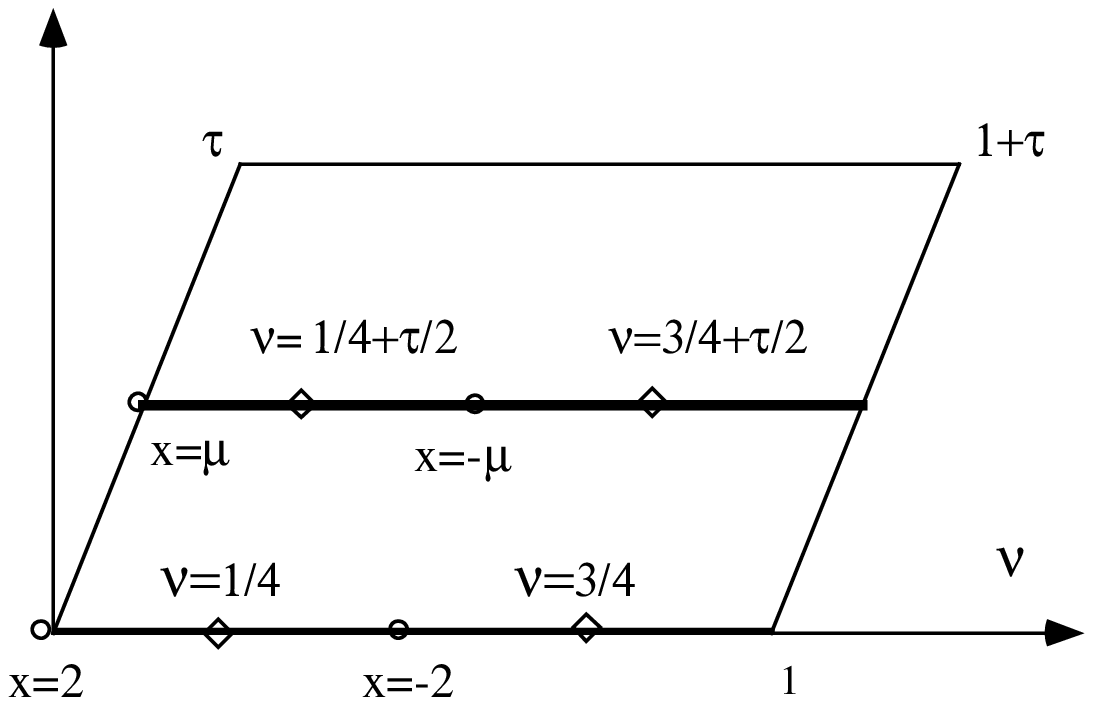, scale=.7}}
\end{center}
\begin{center}
{\footnotesize{\bf Fig. 3:  Diamonds denote the positions
of D4 branes in the $\sp{2k}\times\sp{2k}$ theory}}
\end{center}
\end{figure}

Consider D4 branes at positions $u=a_{i}^{2}$
going around the $x_{6}$ circle in one direction.
These pull the NS5 branes together to meet at $x_{6}=0$.
The branch cut centered on $a_i^2$ in the $u$ plane
(i.e. going around the ``tube'')
is the image under the covering map
of the $x_{10}$ circle at $x_{6}=0$.
The branch points are the images of $x_{10} = 0$ and $\pi R$,
corresponding to $x=2$ and $-2$ respectively.
The D4-brane position $u=a_i^2$ is the image of points
on the $x_{10}$ circle midway between the branch point pre-images,
viz. $x_{10} = \half \pi R$ (or $\threehalf \pi R$),
corresponding to $x=0$.
Therefore $f(u) = 0$ at $u= a_i^2$.

The D4 branes at positions $u=b_{i}^{2}$
going around the $x_{6}$ circle in the other direction
pull the NS5 branes together to meet at $x_{6}=-\pi L$.
The branch cut centered on $b_i^2$ in the $u$ plane
is the image of the $x_{10}$ circle at $x_{6}=-\pi L$,
with the branch point pre-images at
$x_{10} = - \tshalf \theta R$ and $  (\pi-\tshalf \theta) R$
corresponding to $x=\mu$ and $-\mu$ respectively.
The D4-brane position $u=b_i^2$ is the image of
$x_{10} = (\half \pi -\tshalf \theta) R$
(or $(\threehalf \pi -\tshalf\theta) R$),
corresponding to $x=\infty$.
Therefore $f(u) = \infty$ at $u= b_i^2$.

With the poles and zeros of $f(u)$ determined, we may write
the Seiberg-Witten curve as the double cover
\begin{equation}
x = x_{0}\frac{\prod_{i=1}^{k}(u-a^2_{i})}{\prod_{i=1}^{k}(u-b^2_{i})}
\end{equation}
for some $x_0$.
Written as a polynomial in $u$, this becomes
\begin{equation}
u^{k} + \left( A_{1} + \frac{B_{1}}{x-x_{0}} \right) u^{k-1} + \ldots +
\left( A_{k} + \frac{B_{k}}{x-x_{0}}\right) =0
\end{equation}
with $A_{i}$ and $B_{i}$ constants.
The points $(x_0,\eta_0)$ and $(x_0,-\eta_0)$,
where $\eta_0^2 = u (x_0^2-4)(x_0^2-\mu^2)$
correspond to the asymptotic positions of the NS5 branes
as $u \to \infty$.
When rewritten in terms of $v$, this curve agrees exactly with
the results in ref.~\cite{GK}.

\subsection{$\su{N}$ + 2 antisymmetric + 4 fundamental hypermultiplets}

The $\N=2$ SU($N$) gauge theory with hypermultiplets in representations
$2 \Yasymm + 4 \Yfund$ corresponds to a IIA brane configuration
with two NS5 branes, each of them intersecting one of the O$6^{-}$ planes
\cite{U}.
There are a total of $2N$ D4 branes stretched between
the two NS5 branes: $N$ going around the $x_{6}$ circle one way, at
positions $v=a_{i}$, and $N$ going around the $x_{6}$ circle the
other way, at positions $v=-a_{i}$.
In invariant coordinates, there are $N$ mirror pairs of branes at positions
$u=a_{i}^{2}$.  As in sec.~3.2,
the curve we seek is a degree two cover of the
$u$-plane, with a branch cut for each pair of D4 branes.

We can describe a double cover of the $u$ plane with an equation
\begin{equation}
\label{newcover}
\frac{\eta}{x+2}-f(u)=0,
\end{equation}
where $f(u)$ is to be determined,
and $Q$ is described by the elliptic fibration
$ \eta^2 = u (x^2-4)(x-\lambda) $.
For each value of $u$, the left hand side of (\ref{newcover})
is a rational function with poles at $x=-2$ and $x=\infty$,
and two zeroes, which correspond to the sheets of the cover.

Again, consider the universal cover $\C_{v} \times \C_{\nu}$ of $Q$,
choosing $x$ such that
$x=2$, $-2$, $\lambda$, $\infty$
correspond to
$\nu = 0$, $\frac{1}{2}$, $\frac{\tau}{2}$,
and $\frac{1+\tau}{2}$ respectively (see fig.~4).
Then the two sheets of the cover
will be at points $\nu_{1}$ and $\nu_{2}$ in each fiber,
where
$\nu_{1} + \nu_{2} + \frac{1}{2} + \frac{\tau +1}{2} \in \Z  + \Z {\tau}$
(since these are the zeroes and poles of a rational function on $E$).
The sheets coincide
when $2 \nu + \frac{\tau}{2} \in \Z  + \Z {\tau}$, or
$\nu = \fourth \tau $, $\fourth \tau  +\half$,
$\threefourth \tau $, and $\threefourth \tau  +\half$; these
correspond to the branch points of the cover.

\begin{figure}[t]
\begin{center}
\vbox{\epsfig{figure=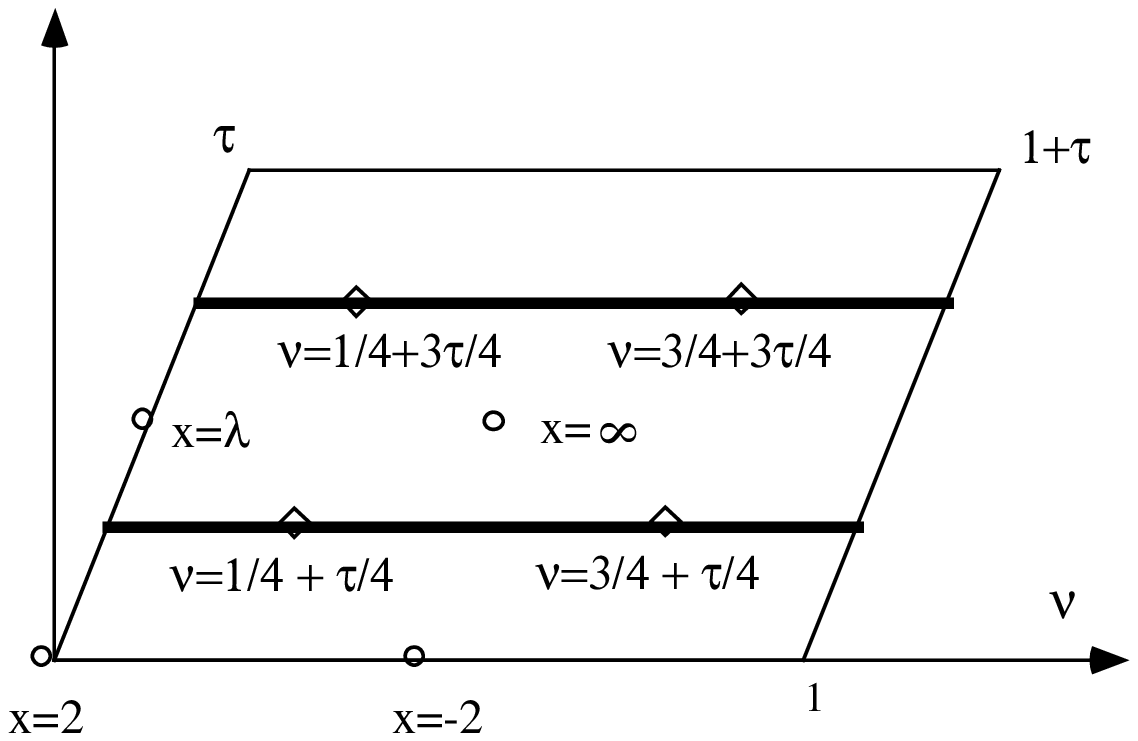, scale=.7}}
\end{center}
\begin{center}
{\footnotesize{\bf Fig. 4: Diamonds denote the positions
of D4 branes in the $\su{N}$ + 2 anti theory}}
\end{center}
\end{figure}

We wish to choose $f(u)$ such that the locations of the D4 branes,
$u=a_{i}^{2}$,
are the images of points $\nu$ midway (along the $x_{10}$ circles)
between the branch point pre-images, namely,
$\nu = \fourth \tau  + \fourth$ and $\fourth \tau  +\threefourth$
(corresponding to $x_6= -\half \pi L$),
and
$\threefourth \tau  + \fourth$ and $\threefourth \tau  +\threefourth$
(corresponding to $x_6= -\threehalf \pi L$).
These points satisfy
$2 \nu + \frac {1}{2} + \frac{\tau}{2} \in \Z  + \Z {\tau}$; in terms
of the cubic curve, each of them is a point of tangency between the curve
and a line through the point corresponding to $\frac {1}{2} +
\frac{\tau}{2}$.  It is easy to check that
these points correspond to points with coordinates $(x,\eta)$
in the quotient surface $Q$ with
$x=2 \pm 2\sqrt{2-\lambda}$, and therefore satisfying
\begin{equation}
\label{positions}
\frac{\eta}{x+2} = \pm \sqrt{u} \sqrt{2-\lambda}.
\end{equation}
The choice of sign in (\ref{positions}) corresponds to the choice
$x_6= -\half \pi L $ or $-\threehalf \pi L$
at which the NS5-branes meet.
This in turn corresponds to choosing the direction in which the D4
branes wrap the $x_{6}$ circle at $v=a_{i}$.
(If the D4 branes wrap one direction at $v=a_{i}$,
then due to the orientifold, they wrap the other direction at $v=-a_{i}$,
so a different choice changes the sign of $v=\sqrt{u}$.)

Thus, comparing (\ref{newcover}) and (\ref{positions}),
we require $f(u)$ to satisfy
\begin{equation}
\label{desirable}
\frac{f(u)^{2}}{u} = 2-\lambda
\qquad \mbox{ when } \qquad  u=a_i^2.
\end{equation}
This can be attained by choosing
\begin{equation}
 f(u) = \frac{ \sqrt{2-\lambda} F_{1}(u)}{F_{2}(u)}
\end{equation}
where $F_1(u)$ and $F_2(u)$  satisfy either
\begin{equation}
\label{firstcondition}
F_{1}(u)^{2} - u F_{2}(u)^{2} = \prod_{i=1}^{N} (u-a_{i}^{2}),
\end{equation}
or,  if $F_{1}(u)$ has a factor of $u$,
\begin{equation}
\label{secondcondition}
\frac{F_{1}(u)^{2}}{u} -  F_{2}(u)^{2} = \prod_{i=1}^{N}
(u-a_{i}^{2}).
\end{equation}

Let us compare this to the results in sec.~5.1 of ref.~\cite{ELNS}.
In the notation of that paper,
\begin{eqnarray}
	 H_{0}(v) = \prod_{i=1}^{N}(v-a_{i}) = \sum_{j=0}^{N} u_{j}v^{N-j},
&\quad& H_{1}(v) = H_{0}(-v) = (-1)^{N} \prod_{i=1}^{N}(v+a_{i})\nonumber\\
	H_{0}(v) = H_{\even}(v) + H_{\odd}(v),
&\quad& H_{1}(v) = (-1)^{N}(H_{\even}(v) - H_{\odd}(v)),\\
	H_{\even} (v) = \sum_{j \ \even} u_{j}v^{N-j},
&\quad& H_{\odd} (v) = \sum_{j \ \odd} u_{j}v^{N-j},\nonumber
\end{eqnarray}
so that $H_{\even}(v)$ is
the even degree part of $H_{0}(v)$ if $N$ is even,
and the odd degree part if $N$ is odd.

If $N$ is even, then $H_{\even}(v)$ and $H_{\odd}(v)/v$
have only even powers of $v$, so can be written as polynomials
$G_{\even}(u)$ and $G_{\odd}(u)$, where $u=v^{2}$.
Then
\begin{equation}
G_{\even}(u)^{2} - u G_{\odd}(u)^{2} =
H^2_{\even}(v) - H^2_{\odd}(v) =
H_{0}(v)H_{1}(v) = \prod_{i=1}^N (u-a_{i}^{2}).
\end{equation}
Since
$ F_{1}(u) = G_{\even}(u)$
and
$ F_{2}(u) = G_{\odd}(u)$ obey condition (\ref{firstcondition}),
we may choose $f(u) = \sqrt{2-\lambda} G_{\even} (u)/ G_{\odd}(u)$.

Similarly, if $N$ is odd,
then $v H_{\even}(v)$ and $H_{\odd}(v)$ have only even powers of $v$,
so we write them also as $G_{\even}(u)$ and $G_{\odd}(u)$.
Then $G_{\even}(u)$ has a factor of $u$, and we have
\begin{equation}
\frac{G_{\even}(u)^{2}}{u} - G_{\odd}(u)^{2} =
H^2_{\even}(v) - H^2_{\odd}(v) =
-H_{0}(v)H_{1}(v) = \prod_{i=1}^N (u-a_{i}^{2}).
\end{equation}
Since $ F_{1}(u) = G_{\even}(u)$
and $ F_{2}(u) = G_{\odd}(u)$ obey condition (\ref{secondcondition}),
we may again choose $f(u) = \sqrt{2-\lambda} G_{\even} (u)/ G_{\odd}(u)$.

The Seiberg-Witten curve for this theory is therefore
\begin{equation}
\label{SWcurve}
\frac{\eta}{x+2} = \sqrt{2-\lambda} \frac{G_{\even}(u)}{G_{\odd}(u)}.
\end{equation}
Observe that,
since $G_{\even}$ has higher degree than $G_{\odd}$,
the right hand side goes to $\infty$ as $u\to\infty$,
thus the asymptotic positions of the NS5-branes are $x=-2$ and $x=\infty$,
corresponding to
$(x_6, x_{10}) = (0, \pi R)$ and
$(-\pi L, (\pi-\tshalf \theta) R)$.

In terms of the double cover $\C_{v} \times E$,
the curve (\ref{SWcurve}) becomes
\begin{equation}
    \label{last equation}
    \frac{y}{x+2} = \sqrt{2-\lambda}
      \frac{\prod_{i=1}^{N}(v-a_{i}) + \prod_{i=1}^{N}(v+a_{i}) }
      {\prod_{i=1}^{N}(v-a_{i}) - \prod_{i=1}^{N}(v+a_{i})}.
\end{equation}
This agrees with the result (5.1.7) of ref.~\cite{ELNS}
and therefore with ref.~\cite{GK},
upon rescaling $u_i$,  which amounts to a redefinition of the $a_i$.

Finally, we observe that, when $v=0$,
the right hand side of eq. (\ref{last equation}) goes to infinity
(and thus $x=-2$ or $\infty$) for even $N$,
whereas it goes to zero
(and thus $x=2$ or $\lambda$) for odd $N$.
That is, when $N$ is even,
the $(x_6,x_{10})$ position of the NS5 branes
(in the $v\to \infty$ limit)
coincides with the $(x_6,x_{10})$ position of those
fixed points of the orientifold (\ref{four fixed points})
through which the curve (\ref{last equation}) passes,
whereas when $N$ is odd,
the $(x_6,x_{10})$ position of the NS5 branes
(in the $v\to \infty$ limit)
coincides with the  $(x_6,x_{10})$ position of those
fixed points through which the curve does {\it not} pass.
This is precisely in accordance with the discussion in section 4.4
of ref.~\cite{U}.

\end{document}